\documentclass[aoas,MSNbibl,nameyear,rotating,dvips]{arximspdf}
\usepackage{multirow}
\usepackage{dcolumn}
\usepackage{graphicx}

\doi{10.1214/11-AOAS493}
\volume{5}
\issue{4}
\pubyear{2011}
\firstpage{2493}
\lastpage{2518}

\makeatletter

\newcolumntype{d}[1]{D{.}{.}{#1}}

\newcommand{\sign}{\operatorname{sign}}
\newcommand{\Poisson}{\operatorname{Poisson}}
\newcommand{\Multinomial}{\operatorname{Multinomial}}
\newcommand{\Gam}{\operatorname{Gamma}}
\newcommand{\Exp}{\operatorname{Exp}}

\newcommand{\Unif}{\operatorname{Unif}}
\newcommand{\NB}{\operatorname{NB}}
\newcommand{\median}{\operatorname{median}}

\makeatother

\begin{document}
\begin{frontmatter}

\title{Classification and clustering of sequencing data using a
Poisson model}
\runtitle{Classification and clustering of sequencing data}

\begin{aug}
\author[A]{\fnms{Daniela M.} \snm{Witten}\corref{}\ead[label=e1]{dwitten@u.washington.edu}}
\runauthor{D. M. Witten}
\affiliation{University of Washington}
\address[A]{Department of Biostatistics\\
University of Washington\\
Box 357232\\
Seattle, Washington 98195-7232\\
USA\\
\printead{e1}} 
\end{aug}

\received{\smonth{9} \syear{2010}}
\revised{\smonth{6} \syear{2011}}

%
\begin{abstract}
In recent years, advances in high throughput sequencing technology have
led to a need for specialized methods for the analysis of digital gene
expression data. While gene expression data measured on a microarray
take on continuous values and can be modeled using the normal
distribution, RNA sequencing data involve nonnegative counts and are
more appropriately modeled using a discrete count distribution, such as
the Poisson or the negative binomial. Consequently, analytic tools
that assume a Gaussian distribution (such as classification methods
based on linear discriminant analysis and clustering methods that use
Euclidean distance) may not perform as well for sequencing data as
methods that are based upon a more appropriate distribution. Here, we
propose new approaches for performing classification and clustering of
observations on the basis of sequencing data. Using a Poisson log
linear model, we develop an analog of diagonal linear discriminant
analysis that is appropriate for sequencing data. We also propose an
approach for clustering sequencing data using a~new dissimilarity
measure that is based upon the Poisson model. We demonstrate the
performances of these approaches in a simulation study, on three
publicly available RNA sequencing data sets, and on a publicly
available chromatin immunoprecipitation sequencing data set.
\end{abstract}

%

\end{frontmatter}

\section{Introduction}
\label{intro}
\subsection{An overview of RNA sequencing data}
\label{introsec}
Since the late 1990s, a vast literature has been devoted to quantifying
the extent to which different
tissue types, biological conditions, and disease states are
characterized by
particular patterns of gene expression, or mRNA levels [examples include
\citet{derisi97}, \citet{Setal98}, \citet{BB99},
\citet{Ramas2002}, \citet{nielsen2002}, \citet{Monti2005}].
During most of that time, the microarray has been the method of
choice for quantifying gene expression.
Though the microarray has led
to an improved understanding of many cellular processes
and disease states, the technology suffers from two fundamental
limitations:
\begin{longlist}[(2)]
\item[(1)] Cross-hybridization can occur, whereby cDNA
hybridizes to a probe for which it is not perfectly matched. This
can lead to high levels of background noise.
\item[(2)] Only transcripts for which a probe is present on the array can
be measured. Therefore, it is not possible to discover novel mRNAs
in a typical microarray experiment.
\end{longlist}

In recent years, \textit{high throughput} or \textit{second generation}
RNA sequencing has emerged as a powerful
alternative to the microarray for measuring gene expression
[see, e.g., \citet{Mortazavi2008}, \citet{Nagalakshmi2008},
\citet{Wilhelm2009}, \citet{Wang09}, \citet{Pepke09}].
This technology allows for the parallel sequencing of a large number
of mRNA transcripts. Briefly, RNA sequencing proceeds as follows
[\citet{Mortazavi2008}, \citet{Morozova09}, \citet
{Wang09}, \citet{Auer2010}, \citet{OshlackReview}]:
\begin{longlist}[(2)]
\item[(1)] RNA is isolated and fragmented to an average length of 200 nucleotides.
\item[(2)] The RNA fragments are converted into cDNA.
\item[(3)] The cDNA is sequenced.
\end{longlist}
This process results in millions of short reads, between 25 and 300
basepairs in length, usually taken from one end of the cDNA fragments
(though some technologies result in ``paired-end'' reads).
The reads are typically
then mapped to the genome or transcriptome if a suitable reference
genome or transcriptome is available; if not, then \textit{de novo}
assembly may be required [\citet{OshlackReview}].
The mapped reads can then be pooled into regions of interest. For
instance, reads
may be pooled by gene or by exon, in which case the data consist of nonnegative
counts indicating the number of reads observed for each gene or each exon.
In this paper we will assume that mapping and pooling of the raw reads
has already been performed. We will
consider RNA data sets that take the form of $n \times p$ matrices,
where $n$ indicates the number of samples for which sequencing was
performed, and $p$ indicates
the number of regions of interest (referred to as ``features''). The
$(i,j)$ element of the data matrix indicates the number of reads
from the $i$th sample that mapped to the $j$th region of interest.
Sequencing data are generally very high dimensional, in the sense that
the number of features $p$ is much larger than the number of
observations $n$. Specifically,~$p$ is usually on the order of tens of
thousands, if not much larger.

RNA sequencing has some major advantages over the microarray. RNA sequencing
data should in theory be much less noisy than microarray data,
since the technology does not suffer from
cross-hybridization. Moreover, novel transcripts and coding regions
can be discovered using RNA sequencing, since unlike studies performed
using microarrays, sequencing experiments do not require
pre-specification of the transcripts of interest. For these reasons,
it seems certain that RNA sequencing is on track to replace the
microarray as the technology of choice for the characterization of
gene expression.

\subsection{Statistical models for RNA sequencing data}
Two aspects of sequencing data are especially worth noting, as they
result in unique statistical challenges.
(1)~Due to artifacts of the sequencing experiment,
different samples can have vastly different total numbers of sequence reads.
This issue is generally addressed by normalizing the samples in some way,
for instance, by
the total number of reads observed for each sample [\citet
{Mortazavi2008}] or a more robust alternative
[\citet{Bullard2010}, \citet{RobinsonOshlack2010},
\citet{AndersHuber10}].
Simply dividing the counts for a given sample by a normalization constant
may not be desirable, since the magnitude of the counts may contain
information about the variability in the data.
(2) Since a sequencing data set consists of the number of reads mapping
to a particular region of interest in a particular sample, the data are
integer-valued and nonnegative. This is in
contrast to microarray data, which is measured on a continuous scale
and can reasonably be modeled using a Gaussian
distribution.

Let $\mathbf X$ denote a $n \times p$ matrix of sequencing data, with $n$
observations (e.g., tissue samples) and $p$ features (regions of
interest; e.g., genes or exons). $X_{ij}$ is the count for feature~$j$
in observation
$i$. For instance, if feature~$j$ is a~gene, then $X_{ij}$
is the total number of reads mapping to gene $j$ in observation~$i$.
A~number of authors have considered a Poisson log linear model for
sequencing data,
%
\begin{equation}\label{mostbasic}
X_{ij} \sim\Poisson(N_{ij}),\qquad N_{ij} = s_i g_j
\end{equation}
[among others, \citet{Marioni08}, \citet{Bullard2010},
\citet{WittenFire2010},
\citet{JunSamSeq}]. To avoid identifiability issues, one can
require $\sum_{i=1}^n s_i = 1$.
This model allows for variability in both the total number of reads per sample
(via the $s_i$ term) and in the total number of reads per region of
interest (via the $g_j$ term).
Since biological replicates seem to be overdispersed relative to the
Poisson model, some
authors have proposed an extension to (\ref{mostbasic}) involving the
use of a negative binomial
model,
a natural alternative to the Poisson model that allows for the variance
to exceed the mean
[among others, \citet{edgeR},\vadjust{\goodbreak} \citet{AndersHuber10}].
Specifically, one could extend
(\ref{mostbasic}) to obtain
%
\begin{equation}\label{negativebinomial}
X_{ij} \sim\mathrm{NB}(N_{ij}, \phi_j),\qquad N_{ij} = s_i g_j,
\end{equation}
where $\mathrm{NB}$ indicates the negative binomial distribution and
$\phi_j \geq0$ is the dispersion parameter for feature $j$.
Throughout this paper, the negative binomial distribution will be
parametrized such that (\ref{negativebinomial}) implies that
observation~$X_{ij}$ has mean $N_{ij}$ and variance $N_{ij} +
N_{ij}^2\phi_j$. When $\phi_j=0$, (\ref{negativebinomial}) reduces
to~(\ref{mostbasic}).

RNA sequencing experiments are often designed such that the $n$
observations are drawn from $K$ different biological conditions, or
\textit{classes}. To accommodate this setting, a number of authors
have extended (\ref{mostbasic}) and (\ref{negativebinomial}) as follows:
%
\begin{eqnarray}
\label{poisson.classes}
X_{ij}| y_i&=&k \sim\Poisson(N_{ij}d_{kj}),\qquad N_{ij} = s_i g_j, \\
\label{nb.classes}
X_{ij}| y_i&=&k \sim\mathrm{NB}(N_{ij}d_{kj}, \phi_j),\qquad N_{ij} = s_i
g_j,
\end{eqnarray}
where $y_i$ indicates the class of the $i$th observation, $y_i \in\{1,
\ldots, K \}$.
Here the $d_{1j}, \ldots, d_{Kj}$ terms allow the $j$th feature to be
differentially expressed between classes. [However, as written in (\ref
{poisson.classes}) and (\ref{nb.classes}), the precise roles of
$d_{1j}, \ldots, d_{Kj}$ are difficult to interpret because the model
is overparametrized.
We will address this point in Section \ref{sectionmod}.]
The models (\ref{poisson.classes}) and (\ref{nb.classes}) have been
used to identify
features that are differentially expressed between conditions
[\citet{Marioni08}, \citet{Bullard2010}, \citet
{WittenFire2010}, \citet{JunSamSeq}, \citet{edgeR},
\citet{AndersHuber10}].\looseness=1

Though the question of how best to identify differentially expressed
features has now been extensively studied, it
is just one of many possible scientific questions that may arise from
sequencing data. This paper addresses the following two
problems:
\begin{longlist}[(2)]
\item[(1)] If each sample is associated with a class label, then one might
wish to build a classifier in order to predict the class label of a
future observation.
\item[(2)] If the samples are unlabeled, one might wish to
cluster the samples in order to identify subgroups among them.
\end{longlist}
Most of the methods in the statistical literature for classification
and clustering implicitly assume a normal distribution for the data. In
this paper, due to the nature of sequencing data,
the Poisson log linear models
(\ref{mostbasic}) and~(\ref{poisson.classes}) will be used to
accomplish these
two tasks.
The importance of the model used can be seen on a simple toy example.
We generated
%
\begin{figure}

\includegraphics{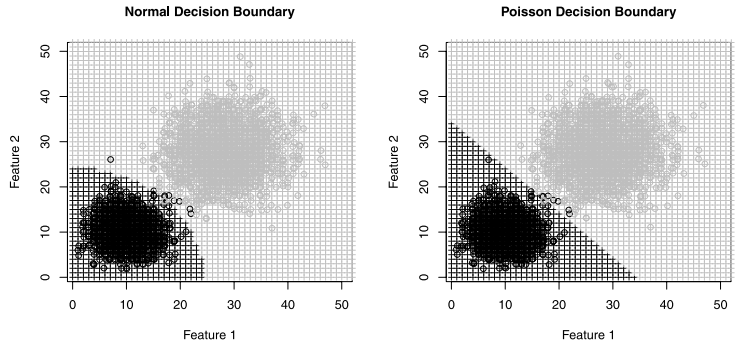}

\caption{Two sets of two-dimensional
independent random variables were generated. The first set of
random variables was generated according to the $\Poisson(10)$
distribution in each dimension, and the second set was generated
according to the $\Poisson(28)$ distribution in each dimension. In
each figure, the two sets of random variables are shown as black
and grey circles, after jittering. The grid in the background of each
plot indicates
the Bayes-optimal decision boundary, assuming a normal distribution
(left) or
a Poisson distribution (right) with the correct mean and variance.}
\label{toyexample2d}
\end{figure}
two-dimensional Poisson distributed random variables with two different
means, each representing a different class. Each dimension was
generated independently. The Bayes-optimal decision boundaries obtained
assuming normality and assuming a Poisson distribution are shown in
Figure \ref{toyexample2d}.

\subsection{Notation and organization}

The following additional notation will be used in this paper.
Let ${\mathbf x}_i = (X_{i1} \cdots X_{ip})^T$ denote row $i$ of $\mathbf X$,
corresponding to the feature measurements for
observation $i$. Also, $X_{\cdot j} = \sum_{i=1}^n X_{ij}$,
$X_{i \cdot} = \sum_{j=1}^p X_{ij}$, $X_{\cdot\cdot} = \sum_{i,j}
X_{ij}$. Moreover, in the classification setting where each
observation belongs to one of $K$ classes, we let $C_k \subset\{ 1,
\ldots, n \}$ contain the indices of the observations in class
$k$---that is, $y_i=k$ if
and only if $i \in C_k$. Furthermore,
$X_{C_k j} = \sum_{i \in C_k} X_{ij}$.

The rest of this paper is organized as follows. In Section
\ref{reviewmodel} the model (\ref{poisson.classes}) is presented in
greater detail, along with methods for fitting it that are based upon
recent proposals in the literature.
This model is used as the basis for the classifier
proposed in Section \ref{proposal.class} and as the basis for the
clustering method proposed in Section~\ref{proposal.clust}. In Section
\ref{simstudy} the performances of the classification and clustering
proposals are evaluated in a simulation
study. The proposals are applied to four sequencing data sets in
Section \ref{realstudy}. Section \ref{disc} contains the Discussion.

\section{A Poisson log linear model for multiple-class sequencing data}
\label{reviewmodel}

\subsection{The model}
\label{sectionmod}
In this paper sequencing data are modeled using a Poisson log linear model.
However, the proposals in this paper could be
extended to the negative binomial model using techniques developed in
\citet{edgeR} and \citet{AndersHuber10}.

The model (\ref{mostbasic}) captures the fact that sequencing data are
characterized by high levels of variation in both the
number of counts per sample ($s_i$) and the number of counts per
feature ($g_j$),
and (\ref{poisson.classes}) additionally allows for the level of
expression of a given feature to depend upon the condition under which
it is observed ($d_{1j}, \ldots,
d_{Kj}$).
Throughout this paper, we assume that the $X_{ij}$'s are independent of
each other for all $i=1, \ldots, n$, $j=1, \ldots, p$.

We first consider the problem of fitting the model (\ref{mostbasic}).
The maximum likelihood estimate (MLE) for $N_{ij}$ is
$\hat N_{ij} = \frac{X_{i \cdot} X_{\cdot j}}{X_{\cdot\cdot}}$
[\citet{Agresti}]. Combining this with the identifiability
constraint that $\sum_{i=1}^n \hat s_i = 1$ yields the estimates
$\hat s_i = X_{i \cdot}/X_{\cdot\cdot}$ and $\hat g_j = {X_{\cdot
j}}$. We can
interpret $\hat s_i$ as an estimate of the \textit{size factor} for
sample $i$, reflecting the fact that different samples may have been
sequenced to different depths. A number of authors have used this size
factor estimate
[\citet{Marioni08}, \citet{Mortazavi2008}].
Recently, it has been pointed out that
$X_{i \cdot}/X_{\cdot\cdot}$ is not a very good estimate for $s_i$
since changes in a few high-count features can have a great effect on
the value of $X_{i \cdot}$, skewing any resulting analyses
[\citet{Bullard2010}, \citet{RobinsonOshlack2010},
\citet{AndersHuber10}, \citet{JunSamSeq}]. For this reason,
several more robust estimates for the size factor $s_i$ have been
proposed. In what follows we will consider three estimates for $s_i$:
\begin{longlist}[(1)]
\item[(1)]\textit{Total count}. We simply use $\hat s_i = {X_{i \cdot
}}/X_{\cdot\cdot}$, the total count for the $i$th observation, which
is based upon the MLE for $N_{ij}$
under the model (\ref{mostbasic}).
\item[(2)]\textit{Median ratio}. \citet{AndersHuber10} propose the use of
$\hat s_i = m_i /\break\sum_{i=1}^n m_i$, where
%
\begin{equation}\label{mr}
m_i = \median_j \biggl\{ \frac{X_{ij}}{(\prod_{i'=1}^n X_{i'j})^{1/n}} \biggr\}.
\end{equation}
That is, the size factor for the $i$th sample is obtained by computing
the median, over all $p$ features, of the $i$th sample count for that
feature divided by the geometric mean of
all sample counts for that feature.
%
\item[(3)]\textit{Quantile}. \citet{Bullard2010} propose taking $\hat
s_i = q_i / \sum_{i=1}^n q_i$, where~$q_i$ is the 75th percentile of
the counts for each sample.
\end{longlist}
Thus, throughout this paper, we will estimate $N_{ij}$ in (\ref
{mostbasic}) according to $\hat N_{ij} = \hat s_i \hat g_j$, where
$\hat s_i$ is given by one of the methods described above,
and
$\hat g_j = {X_{\cdot j}}$.

We now consider the problem of fitting the model (\ref
{poisson.classes}). Since we would like to attribute as much as
possible of the observed variability in the counts for each feature to
sample and feature effects ($s_i$ and $g_j$) rather than to class
differences~($d_{kj}$), we estimate $N_{ij}$ under the model (\ref
{mostbasic}) without making use of the class labels.
We then estimate $d_{kj}$ by treating $\hat N_{ij}$ as an offset in the
model (\ref{poisson.classes}). That is, we fit the model
%
\begin{equation}\label{model}
X_{ij} | y_i = k \sim\Poisson(\hat N_{ij} d_{kj}).
\end{equation}
Maximum likelihood provides a natural way to estimate $d_{kj}$ in (\ref
{model}),
yielding
$\hat d_{kj} = \frac{ X_{C_k
j}}{\sum_{i \in C_k} \hat N_{ij}}$.
Now\vspace*{-3pt} $\hat d_{kj}$ has a simple interpretation: if $\hat d_{kj}>1$, then
the $j$th feature is over-expressed relative to the baseline in the
$k$th class, and if $\hat d_{kj}<1$,
then the $j$th feature is under-expressed relative to the baseline in
the $k$th class.

However, if $X_{C_k j}=0$ (an event that is not unlikely if the true
mean for feature $j$ is small), then the maximum likelihood
estimate for $d_{kj}$ equals zero.
This can pose a problem for downstream analyses, since this estimate
completely precludes the possibility of a nonzero count for feature~$j$ arising from an observation in class~$k$.
We can remedy this problem by putting a $\Gam(\beta, \beta)$ prior
on $d_{kj}$ in
the model (\ref{model}). Here,
the shape and rate parameters both equal $\beta$.
Then, the posterior distribution for $d_{kj}$ is
$\Gam(X_{C_k j} + \beta, \sum_{i \in C_k} \hat N_{ij} +\beta)$, and
the posterior mean is
%
\begin{equation}\label{postmean}
\hat d_{kj} = \frac{ X_{C_k
j} + \beta}{ \sum_{i \in C_k} \hat N_{ij} +
\beta}.
\end{equation}
Equation (\ref{postmean}) is a smoothed estimate of
$d_{kj}$ that behaves well even if $X_{C_k j}=0$ for some class $k$. We
took $\beta=1$ in all of the examples shown in this paper.

\subsection{A transformation for overdispersed data}
\label{overdisp}
A number of authors have observed that, in practice, biological
replicates of sequencing data
tend to be overdispersed relative to the Poisson model, in the sense
that the variance is larger
than the mean. This problem could be
addressed by using a~different model for the data, such as a negative
binomial model [\citet{edgeR}, \citet{AndersHuber10}].
Instead, we apply a power transformation
to the data [\citet{WittenFire2010}, \citet{JunSamSeq}].
The transformation $X_{ij}'
\leftarrow X_{ij}^{\alpha}$ is used, where
$\alpha\in(0,1]$ is chosen so that
%
\begin{equation} \label{transform}
\sum_{i=1}^n \sum_{j=1}^p \frac{(X_{ij}' - {X_{i \cdot}'
X_{\cdot
j}'}/{X_{\cdot\cdot}'})^2}{{X_{i \cdot}' X_{\cdot
j}'}/{X_{\cdot\cdot}'}} \approx(n-1)(p-1).
\end{equation}
This is simply a test of the goodness of fit of the model (\ref
{mostbasic}) to the data [\citet{Agresti}], using the total count
size factor estimate $\hat s_i = X_{i \cdot}'/X_{\cdot\cdot}'$.

Though the resulting transformed data are not integer-valued, we
nonetheless model them using the Poisson distribution.
This simple transformation allows us to use a Poisson model even in the
case of overdispersed data, in order to avoid having to fit a
necessarily more complicated negative binomial model
(a task made especially complicated in the typical setting for
sequencing data where the number of samples is small). In Section \ref
{simstudy} we show that even when data are generated according to
a negative binomial model with moderate overdispersion, the
classification and clustering proposals based on the Poisson model
perform well on the transformed data.

\section{A proposal for classifying sequencing data}
\label{proposal.class}

\subsection{The Poisson linear discriminant analysis classifier}
Suppose that we wish to classify a test observation ${\mathbf x}^* = (X_1^*
\cdots X_p^*)^T$ on the basis\vadjust{\goodbreak} of training data $\{ ({\mathbf x}_i, {y}_i) \}
_{i=1}^n$. Let $y^*$ denote the unknown class label.
By Bayes' rule,
%
\begin{equation}\label{classifrule}
P(y^*=k|{\mathbf x}^*) \propto f_k({\mathbf x}^*) \pi_k,
\end{equation}
where $f_k$ is the density of an observation in class $k$ and $\pi_k$
is the prior probability that an observation belongs to class $k$. If
$f_k$ is a normal density with
a class-specific mean and
common covariance, then assigning an observation to the class for which
(\ref{classifrule}) is largest results in standard LDA
[for a reference, see \citet{ElemStatLearn}]. If we instead assume
that the observations are normally distributed with a class-specific
mean and a common diagonal covariance matrix, then diagonal LDA
results [\citet{dudoit2001}]. The assumption of
normality is not appropriate for sequencing data, and neither is the
assumption of a common covariance matrix for the $K$ classes. We
instead assume that the
data arise from the model (\ref{poisson.classes}), and we also assume
that the features are independent. The assumption of independence is
often made for high-dimensional continuous data
[e.g., see \citet{dudoit2001}, Tibshirani et~al.
(\citeyear{THNC2002,THNC2003}), \citet{BickelLevina04}, \citet
{WittenTibsPenLDA2011}] since when $p>n$, there are too few observations
available to be able to effectively estimate the dependence
structure among the features.

Evaluating (\ref{classifrule}) requires estimation of $f_k({\mathbf
x}^*)$ and $\pi_k$. The model (\ref{poisson.classes}) states that
$X_j^* | y^* = k \sim\Poisson(s^* g_j d_{kj})$. We first estimate $s_1,
\ldots, s_n$, the size factors for the training data, using the total
count, quantile, or median ratio approaches (Section~\ref{sectionmod}).
We then estimate $g_j$ and $d_{kj}$ by evaluating $\hat g_j = X_{\cdot
j}$ and (\ref{postmean}) on the training data. Finally, we estimate
$s^*$ as follows:
\begin{itemize}
\item If the total count estimate for the size factors was used, then
$\hat s^* = \sum_{j=1}^p X_j^*/\break X_{\cdot\cdot}$, where $X_{\cdot
\cdot}$ is the total number of counts on the \textit{training} data.
\item If the median ratio estimate for the size factors was used, then
$\hat s^* = m^* / \sum_{i=1}^n m_i$, where
$m^* = \median_j \{ \frac{X_j^*}{(\prod_{i=1}^n X_{ij})^{1/n}} \}
$---note that the denominator is
the geometric mean for the $j$th feature among the \textit{training}
observations. Here $m_i$ is given by~(\ref{mr}).
\item If the quantile estimate for the size factors was used, then
$\hat s^* = q^* / \sum_{i=1}^n q_i$. Here, $q^*$ is the 75th
percentile of counts for the test observation,
and $q_i$ is the 75th percentile of counts for the $i$th training observation.
\end{itemize}
Note that these estimates of $s^*$ are the direct extensions of the
size factor estimates presented in Section \ref{sectionmod}, applied
to the test observation ${\mathbf x}^*$.

We now consider the problem of estimating $\pi_k$. We could let
$\hat\pi_1 = \cdots= \hat\pi_K = 1/K$, corresponding to the prior
that all classes are equally likely. Alternatively, we could let $\hat
\pi_k = {|C_k|}/{n}$, if we believe that the proportion of
observations in each class seen in the training set is representative
of the proportion in the population. In the examples presented in
Sections \ref{simstudy} and~\ref{realstudy}, we take $\hat\pi_1 =
\cdots= \hat\pi_K = 1/K$.

Plugging these estimates into (\ref{poisson.classes}) and recalling
our assumption of independent features, (\ref{classifrule}) yields
%
\begin{eqnarray}\label{classifruleexplicit}
\log{\widehat{P(y^*=k | {\mathbf x}^*)}} & = & \log{\hat{f}_k({\mathbf x}^*)}
+ \log\hat\pi_k +
c \nonumber\\[-8pt]\\[-8pt]
& = &\sum_{j=1}^p X_{j}^* \log\hat d_{kj} - \hat s^* \sum_{j=1}^p \hat g_j
\hat d_{kj} + \log\hat\pi_k +c',
\nonumber
\end{eqnarray}
where $c$ and $c'$ are constants that do not depend on the class
label. Only the first term in (\ref{classifruleexplicit}) involves the
individual feature measurements
for the test observation~${\mathbf x}^*$. Therefore, the
classification rule that assigns the test observation
to the class for which (\ref{classifruleexplicit}) is largest is
linear in ${\mathbf x}^*$. For this reason, we call this classifier
\textit{Poisson linear discriminant analysis} (PLDA). This name reflects
the linearity of the classifier, as well as the fact that it differs
from standard LDA only in its use of a Poisson model for the data.

\subsection{The sparse PLDA classifier}
PLDA's classification rule (\ref{classifruleexplicit}) is quite
simple, in that it is linear in the components of ${\mathbf x}^*$. But when
the estimate~(\ref{postmean}) is used for $d_{kj}$,
then $\hat d_{kj} \neq1$ in general and so the classification rule
(\ref{classifruleexplicit}) involves all $p$ features. For sequencing
data, $p$ may be quite large, and a~classifier that involves only a
subset of the features is desirable in order to achieve increased
interpretability and reduced variance. By inspection, the
classification rule (\ref{classifruleexplicit})
will not involve the data for feature $j$ if
$\hat d_{1j} = \cdots=\hat d_{Kj}=1$.
We obtain a classification rule that is sparse in the features by using
the following estimate for $d_{kj}$ in (\ref{classifruleexplicit}),
which shrinks the estimate (\ref{postmean}) toward 1:
%
\begin{equation}\label{sparseest}
\hat{d}_{kj} =
\cases{\displaystyle \frac{a}{b}-\frac{\rho}{\sqrt{b}}, &\quad if
$\displaystyle \sqrt{b}\biggl(\frac{a}{b}-1\biggr) >\rho$,\vspace*{2pt}\cr
\displaystyle \frac{a}{b}+\frac{\rho}{\sqrt{b}}, &\quad if
$\displaystyle \sqrt{b}\biggl(1-\frac{a}{b}\biggr)
>\rho$,\vspace*{2pt}\cr
1, &\quad if $\displaystyle \sqrt{b}\biggl|1-\frac{a}{b}\biggr| <\rho$,}
\end{equation}
where $a=X_{C_k j} + \beta$ and $b=\sum_{i \in C_k} \hat N_{ij} +
\beta$. Here, $\rho$ is a nonnegative
tuning parameter that is generally chosen by cross-validation. When
$\rho=0$, (\ref{sparseest}) is simply the estimate~(\ref{postmean}).
As $\rho$ increases, so does the number of estimates~(\ref
{sparseest}) that are exactly equal to 1. Using the estimate~(\ref
{sparseest}) in the PLDA classifier~(\ref{classifruleexplicit}) yields
\textit{sparse PLDA} (\textit{sPLDA}).
The operation (\ref{sparseest}) can be written more concisely as $\hat
d_{kj} = 1+S(a/b-1, \rho/\sqrt{b})$, where $S$ is the
soft-thresholding operator, given by $S(x,a)=\sign(x)(|x|-a)_+$. Note
that the form of (\ref{sparseest}) combined with the definition of $b$
implies that if class $k$ contains few observations, or if the mean for
class $k$ in feature $j$ is small, then the estimate for~$d_{kj}$ will
undergo greater shrinkage.

Sparse PLDA is closely related to the nearest shrunken centroids (NSC)
classifier [Tibshirani et~al.
(\citeyear{THNC2002,THNC2003})], which is a
variant of diagonal LDA that arises from
shrinking the class-specific mean vectors toward a common mean using
the soft-thresholding operator.
In fact, sPLDA arises from replacing the normal model that leads to NSC
with the Poisson model (\ref{poisson.classes}).
For this reason, NSC is a natural method against which to compare sPLDA.

\section{A proposal for clustering sequencing data}
\label{proposal.clust}
\subsection{Poisson dissimilarity}
\label{proposal.clust.deets}
We now consider the problem of computing a $n \times n$ dissimilarity
matrix for $n$ observations for which sequencing measurements are
available. For microarray data,
squared Euclidean distance is a common choice of dissimilarity measure.
Another popular choice,
correlation-based distance, is equivalent to squared Euclidean distance
up to a scaling of the observations [\citet{ElemStatLearn}].
Squared Euclidean distance can be derived as the consequence of
performing hypothesis testing on a simple Gaussian model for the
data. That is, consider the model
%
\begin{equation}\label{normalmodel}
X_{ij} \sim N(\mu_{ij}, \sigma^2),\qquad X_{i'j} \sim N(\mu_{i'j},
\sigma^2),
\end{equation}
where we assume that the features and observations
are independent. Consider testing the null hypothesis $H_0\dvtx \mu
_{ij}=\mu_{i'j}, j=1, \ldots, p$, against $H_a$, which states that
$\mu_{ij}$ and $\mu_{i'j}$ are
unrestricted. The resulting log likelihood ratio statistic is
proportional to
%
\begin{eqnarray}\qquad
\sum_{j=1}^p \biggl( X_{ij} - \frac{X_{ij}+X_{i'j}}{2} \biggr)^2 + \sum_{j=1}^p
\biggl( X_{i'j} - \frac{X_{ij}+X_{i'j}}{2} \biggr)^2 &\propto&\sum_{j=1}^p (
X_{ij} - X_{i'j} )^2 \nonumber\\[-8pt]\\[-8pt]
&=& \|{\mathbf x}_i - {\mathbf x}_{i'}\|^2.\nonumber
\end{eqnarray}
Therefore, squared Euclidean distance is equivalent to a log likelihood ratio
statistic for each pair of observations, under a Gaussian model for the
data.\looseness=1

Now, as discussed earlier, the model (\ref{normalmodel}) does not seem
appropriate for
sequencing data.
Instead, consider the model
%
\begin{eqnarray}\label{poimodclust}
X_{ij} &\sim&\Poisson(N_{ij} d_{ij}),\qquad
X_{i'j} \sim\Poisson(N_{i'j}
d_{i'j}),\nonumber\\[-8pt]\\[-8pt]
N_{ij} &=& s_i g_j,\qquad N_{i'j} = s_{i'} g_j,\nonumber
\end{eqnarray}
where we assume that the features are independent.
This is simply the
model~(\ref{poisson.classes}), restricted to
${\mathbf x}_i$ and ${\mathbf x}_{i'}$. We first estimate
$N_{ij}, N_{i'j}$ under the simpler model
%
\begin{eqnarray}\label{poimodclust.simple}
X_{ij} &\sim&\Poisson(N_{ij} ),\qquad X_{i'j} \sim\Poisson(N_{i'j}),\nonumber\\[-8pt]\\[-8pt]
N_{ij} &=& s_i g_j,\qquad N_{i'j} = s_{i'} g_j\nonumber
\end{eqnarray}
as described in Section \ref{sectionmod}---using total count,
quantile, or median ratio size factor estimates---but restricted to
${\mathbf x}_i$ and ${\mathbf x}_{i'}$.
We then test the null hypothesis
$H_0\dvtx d_{ij} = d_{i'j}=1, j=1,\ldots,p$, against the alternative $H_a$,
which states
that $d_{ij}$ and $d_{i'j}$ are nonnegative. The resulting log likelihood
ratio statistic can be used as a measure of dissimilarity between
${\mathbf x}_i$ and ${\mathbf x}_{i'}$. A~standard log likelihood ratio
statistic would involve computing the maximum likelihood estimates for
$d_{ij}$ and $d_{i'j}$ under $H_a$. However, to avoid the estimate
$\hat d_{ij}=0$ if $X_{ij}=0$, we instead compute a \textit{modified}
log likelihood ratio statistic: we evaluate the log likelihood under
$H_a$ using the estimates
%
\begin{equation}\label{part2}
\hat d_{ij} = \frac{X_{ij} + \beta}{\hat N_{ij}+\beta},\qquad \hat d_{i'j}
= \frac{X_{i'j} + \beta}{\hat N_{i'j}+\beta},
\end{equation}
which are the posterior means for $d_{ij}$ and $d_{i'j}$ under $\Gam
(\beta,\beta)$ priors.
The resulting modified log likelihood ratio statistic is
%
\begin{equation}\label{loglrpoi}
\sum_{j=1}^p (\hat N_{ij} + \hat N_{i'j} - \hat N_{ij} \hat d_{ij} -
\hat N_{i'j} \hat d_{i'j} + X_{ij} \log\hat d_{ij} + X_{i'j} \log
\hat d_{i'j}).
\end{equation}
Then (\ref{loglrpoi}) can be thought of as the dissimilarity between
${\mathbf x}_i$ and ${\mathbf x}_{i'}$ under the
model~(\ref{poisson.classes}). The dissimilarity between two identical
observations is 0, and all dissimilarities are nonnegative (see
the \hyperref[app]{Appendix}). We will refer to the $n \times n$ dissimilarity matrix with
$(i,j)$ element given by (\ref{loglrpoi}) as the \textit {Poisson
dissimilarity} matrix.

Hierarchical clustering is a very popular approach for clustering in
genomics since it leads to a visual representation of the data and does
not require prespecifying the number of clusters.
Hierarchical clustering operates on a $n \times n$ dissimilarity
matrix, and can be performed using
the Poisson dissimilarity matrix. We will refer to the clustering
obtained using this dissimilarity matrix as \textit{Poisson clustering}.

To obtain a $p \times p$ Poisson dissimilarity matrix of the features
rather than a~$n \times n$ dissimilarity matrix of the observations,
one could use a Poisson model and
repeat the arguments in this section,
reversing the roles of observations and features in each of the
relevant equations. One could then use this dissimilarity matrix in
order to perform Poisson clustering of the features.


\subsection{Alternative approaches for clustering count data}
\label{sec:cluster}
We briefly review some approaches from the literature for computing
dissimilarity matrices or performing clustering using count data.

Serial analysis of gene expression (SAGE) is a sequencing-based
method for gene expression profiling that predates RNA sequencing
[\citet{SAGEWang07}]. In SAGE a prespecified region of the RNA
transcript is sequenced,
and so the ability to detect previously unknown RNA transcripts is
somewhat limited. \citet{Cai04} propose a procedure for
performing $K$-means clustering of SAGE
data using a Poisson model. Though their approach focuses on clustering
features (known as \textit{tags} in the context of SAGE data) rather
than observations,
their approach is fundamentally very similar to the one proposed here
in that a Poisson model is used and deviations from the Poisson model
(measured using a chi-squared
statistic or the log likelihood) are taken as an indication that two
tags are different from each other and hence belong in different
clusters. They propose to fit the Poisson model using the maximum
likelihood parameter estimates, which is analogous to using total count
size factor estimates in our model. They fit the Poisson model using
all $n$ observations at once rather than separately
for each pair of observations as in~(\ref{poimodclust.simple}). In our
experience, fitting the model separately for each pair of observations
leads to better results.
Their approach yields a~prespecified number of clusters $K$, whereas
ours yields a dissimilarity matrix that can be hierarchically clustered
to obtain
any number of clusters, or used for other purposes.

\citet{Zavolan08} propose a method for
computing a dissimilarity matrix using sequencing data that is also
very closely related to ours. They assume that each
observation is drawn from a multinomial distribution, and they test
whether or not the multinomial parameters for each pair of
observations are equal. This is almost identical to our Poisson model
and associated hypothesis testing framework, since if the observations
are distributed according to (\ref{poimodclust}), then their
distribution conditional on $X_{i \cdot}$, $X_{i' \cdot}$ is
multinomial. In fact, the log likelihood ratio statistics under our
model and theirs are identical for certain very natural estimates
of $N_{ij}$, $N_{i'j}$, $d_{ij}$, and $d_{i'j}$ in (\ref{loglrpoi})
(see the \hyperref[app]{Appendix}). However, there are some important
differences between the two proposals. \citet{Zavolan08} place a
Dirichlet prior on the parameters for
the multinomial distribution, and then use a Bayes factor as a measure
of the dissimilarity between two observations.
Consequently, two identical observations can have nonzero dissimilarity
according to \citet{Zavolan08}, and two different
observations can have smaller dissimilarity than two identical
observations. This leads to problems in the interpretation of their
dissimilarity measure as well as in the performance of any clustering
approach that is based upon it.
Finally, their approach
can suffer from numerical issues where
the computed dissimilarity between a pair of observations rounds to zero.

The \texttt{edgeR} software package, available from \texttt{Bioconductor}
[\citet{edgeR}], provides a tool for measuring the dissimilarity
between a pair of observations
based upon a negative
binomial model. The $q$ features with highest feature-wise dispersion
across all $n$ samples are selected, and the common dispersion of those
$q$ features
for each pair of observations is used
as a measure of pairwise dissimilarity.

Finally, \citet{AndersHuber10} propose a variance-stabilizing
transformation based on the negative binomial model, and suggest
performing standard clustering procedures
on the transformed data---for instance, one could perform hierarchical
clustering after computing the squared Euclidean distances between the
transformed observations.

%
\begin{table}
\caption{Summary of approaches for computing
dissimilarity measures}\label{tab:cluster}
\begin{tabular*}{\tablewidth}{@{\extracolsep{\fill}}lp{240pt}@{}}
\hline
\textbf{Method}&\multicolumn{1}{c@{}}{\textbf{Description}}\\
\hline
EdgeR& A proposal in the edgeR software package, based on
a~negative binomial model [\citet{edgeR}]. The dissimilarity
between a pair of observations is
computed as the
common dispersion of the 500 features with the highest feature-wise
dispersion across all $n$ samples.\\
[4pt]
Berninger& An approach for computing dissimilarities
between pairs of observations, using a multinomial model with a
Dirichlet prior [\citet{Zavolan08}].\\
[4pt]
VST& A variance stabilizing transformation (VST) based on
a negative binomial model [\citet{AndersHuber10}] is applied to
the data,
and then squared Euclidean distances between
pairs of observations are computed.\\
[4pt]
Sq. Euclidean total count& Squared Euclidean distances
are computed after scaling each sample by the total count size factor
estimate (Section \ref{sectionmod}).\\
[4pt]
Sq. Euclidean quantile& Squared Euclidean distances are
computed after scaling each sample by the quantile size factor estimate
[Section \ref{sectionmod}; \citet{Bullard2010}].\\
[4pt]
Sq. Euclidean median ratio& Squared Euclidean distances
are computed after scaling each sample by the median ratio size factor
estimate [Section \ref{sectionmod}; \citet{AndersHuber10}].\\
[4pt]
Poisson total count& The data are transformed as in
Section \ref{overdisp}. Then Poisson dissimilarity is computed
according to (\ref{loglrpoi}) using the total count size factor
estimate (Section \ref{sectionmod}).\\
[4pt]
Poisson quantile& The data are transformed as in Section
\ref{overdisp}. Then Poisson dissimilarity is computed
according to (\ref{loglrpoi}) using the quantile size factor estimate
[Section \ref{sectionmod}; \citet{Bullard2010}].\\
[4pt]
Poisson median ratio& The data are transformed as in
Section \ref{overdisp}. Then Poisson dissimilarity is computed
according to (\ref{loglrpoi})
using the median ratio size factor estimate [Section \ref{sectionmod};
\citet{AndersHuber10}].\\
\hline
\end{tabular*}
\end{table}

In Sections \ref{simstudy} and \ref{realstudy} we compare our
clustering proposal to these competing approaches. Details of the
approaches used in the comparisons are given
in Table~\ref{tab:cluster}.

\section{A simulation study}
\label{simstudy}

\subsection{Simulation setup}
\label{simstudy.setup}

We generated data under the model
%
\begin{equation}\label{NBMod}
X_{ij} | y_i = k \sim\NB(s_i g_j d_{kj}, \phi),
\end{equation}
where $\phi$ is the dispersion parameter for the negative binomial
distribution. Therefore, given that $y_i=k$, $X_{ij}$ has mean $s_i g_j
d_{kj}$ and variance
$s_i g_j d_{kj} + (s_i g_j d_{kj})^2 \phi$. We tried three values of
$\phi\dvtx\phi=0.01$ (very slight overdispersion),
$\phi=0.1$ (substantial overdispersion), and $\phi=1$ (very high
overdispersion).
There were $K=3$ classes.
The size factors are independent and identically distributed,
$s_i \sim\Unif(0.2, 2.2)$. The $g_j$'s are independent and
identically distributed,
$g_j \sim\Exp(1/25)$.
Each of the $p=10\mbox{,}000$ features had a $30\%$ chance of being
differentially expressed between classes. If a feature was not
differentially expressed, then $d_{1j} = d_{2j} = d_{3j} = 1$. If a
feature\vspace*{2pt} was differentially expressed, then $\log d_{kj} = z_{kj}$ where
the $z_{kj}$'s are independent and identically distributed,
%
\begin{sidewaystable}
\textwidth=\textheight
\tablewidth=\textwidth
\caption{Simulation results: nine classification methods.
NSC, NSC on $\sqrt{X_{ij}+3/8}$, and sPLDA were performed, using three
different size factor estimates:
total count (TC), quantile (Q), and median ratio (MR).
Cross-validation was performed on a training set of $n$ observations,
and error rates were computed on $n$ test observations.
We report the mean numbers of test errors and nonzero features over 50
simulated data sets. Standard errors are in parentheses}
\label{tab:simclass}
\begin{tabular*}{\tablewidth}{@{\extracolsep{\fill}}ld{1.2}d{1.3}ccccccc@{}}
\hline
$\bolds{n}$&\multicolumn{1}{c}{$\bolds{\phi}$} &\multicolumn{1}{c}{$\bolds{\sigma}$}
& \multicolumn{1}{c}{\textbf{Method}}
& \multicolumn{1}{c}{\textbf{NSC err.}}
& \multicolumn{1}{c}{\textbf{NSC sqrt err.}}
& \multicolumn{1}{c}{\textbf{sPLDA err.}}
& \multicolumn{1}{c}{\textbf{NSC nonzero}}
& \multicolumn{1}{c}{\textbf{NSC sqrt nonzero}}
& \multicolumn{1}{c@{}}{\textbf{sPLDA nonzero}} \\
\hline
12&0.01&0.05&TC &
\phantom{0}4.18 (0.34) & \phantom{0}5.74 (0.28) & \phantom{0}2.24 (0.26) & 1947.6 (441.3) & 2217.9 (509.0) &
\hphantom{0}791.4 (111.7) \\
&&&Q & \phantom{0}4.38 (0.34) & \phantom{0}5.82 (0.26) & \phantom{0}2.26 (0.25) & 1670.6 (394.5) &
2010.1 (478.2) & \hphantom{0}782.3 (110.0) \\
&&&MR & \phantom{0}4.28 (0.34) & \phantom{0}5.78 (0.27) & \phantom{0}2.20 (0.24) & 1731.8 (402.6) &
2327.8 (517.9) & \hphantom{0}795.4 (110.8) \\
[4pt]
50&0.01&0.025&TC &
19.14 (0.67) & 24.06 (0.70) & 16.84 (0.55) & 2316.6 (398.8) & 3122.5 (516.6)
& 1830.7 (217.9) \\
&&&Q & 20.32 (0.71) & 24.82 (0.70) & 17.14 (0.56) & 1870.7 (335.2) &
3380.9 (519.4) & 1860.2 (229.6) \\
&&&MR & 19.66 (0.69) & 24.48 (0.69) & 16.88 (0.60) & 2488.7 (437.8) &
2698.7 (513.4) & 1934.9 (224.5) \\
[4pt]
12&0.1&0.1&TC &
\phantom{0}2.52 (0.31) & \phantom{0}2.66 (0.26) & \phantom{0}1.58 (0.25) & 5143.2 (527.2) & 2738.5 (461.2) &
3878.2 (369.4) \\
&&&Q & \phantom{0}2.12 (0.27) & \phantom{0}2.68 (0.26) & \phantom{0}1.62 (0.26) & 5207.0 (536.8) &
2879.8 (456.7) & 3927.2 (371.7) \\
&&&MR & \phantom{0}2.28 (0.29) & \phantom{0}2.88 (0.28) & \phantom{0}1.60 (0.26) & 4849.4 (531.2) &
2932.0 (477.3) & 3889.2 (368.6) \\
[4pt]
50&0.1&0.05&TC &
16.80 (0.54) & 17.76 (0.61) & 17.94 (0.70) & 3802.5 (408.1) & 3785.5 (418.1)
& 3308.5 (355.1) \\
&&&Q & 17.08 (0.64) & 17.16 (0.60) & 17.88 (0.65) & 4293.2 (479.1) &
3921.5 (371.8) & 3284.0 (352.8) \\
&&&MR & 16.78 (0.59) & 17.34 (0.66) & 17.96 (0.71) & 3475.3 (392.4) &
4398.3 (457.0) & 3489.3 (371.9) \\
[4pt]
12&1&0.2&TC &
\phantom{0}3.24 (0.25) & \phantom{0}4.28 (0.35) & \phantom{0}4.26 (0.32) & 8846.2 (380.7) & 6127.8 (524.6) &
4502.0 (509.9) \\
&&&Q & \phantom{0}3.20 (0.23) & \phantom{0}4.04 (0.33) & \phantom{0}4.08 (0.29) & 8991.8 (318.8) &
6342.1 (557.6) & 4551.7 (512.1) \\
&&&MR & \phantom{0}3.22 (0.26) & \phantom{0}3.60 (0.30) & \phantom{0}4.00 (0.31) & 8389.0 (396.4) &
7082.7 (515.5) & 4518.9 (514.6) \\
[4pt]
50&1&0.1&TC &
25.56 (0.61) & 25.80 (0.55) & 25.66 (0.50) & 4237.8 (503.5) & 4293.5 (495.9)
& 3150.5 (433.0) \\
&&&Q & 25.82 (0.61) & 25.90 (0.64) & 26.02 (0.55) & 4629.1 (516.2) &
4170.5 (491.7) & 3131.2 (406.6) \\
&&&MR & 25.92 (0.68) & 25.86 (0.59) & 25.52 (0.51) & 4427.5 (524.0) &
4362.6 (498.0) & 3156.8 (410.4) \\
\hline
\end{tabular*}
\end{sidewaystable}
$z_{kj} \sim N(0, \sigma^2)$. The value of $\sigma$ depended on the
simulation and is specified in Tables \ref{tab:simclass} and \ref
{tab:simclust} below.

\subsection{Evaluation of sparse PLDA}
\label{simstudy.class}

We considered three classifiers:
\begin{longlist}[(2)]
\item[(1)] NSC [Tibshirani et~al.
(\citeyear{THNC2002,THNC2003})] after
dividing each observation by a size factor estimate.
\item[(2)] NSC after dividing each observation by a size factor estimate,
and then transforming as follows: $X_{ij}' \leftarrow\sqrt{X_{ij} +
3/8}$. This transformation should make Poisson random variables have
approximately constant variance
[Ans\-combe (\citeyear{Anscombe1948})].
\item[(3)] sPLDA after performing the power transformation described in
Section \ref{overdisp}.
\end{longlist}
For each classifier, three size factor estimates were used: total
count, quantile, and median ratio. These are described in Section \ref
{sectionmod}.
Therefore, a total of nine classification methods were considered.
Results are shown in Table~\ref{tab:simclass}. sPLDA performs quite
well when there is little overdispersion relative to the Poisson
model---that is, when the dispersion parameter, $\phi$, in the model~%
(\ref{NBMod}) is small. The performance of sPLDA relative to NSC
deteriorates when the data are very overdispersed
relative to the Poisson model. Moreover, the square root transformation
on the whole seemed to lead to substantially worse results for the NSC
classifier.

%
\begin{table}
\caption{Simulation results: ten clustering methods. Complete-linkage
hierarchical clustering was applied to the nine dissimilarity measures
described in Table \protect\ref{tab:cluster}, and each dendrogram was
cut in order to obtain three clusters. The proposal of Cai et al.
(\protect\citeyear{Cai04}) is also included; this required specifying
$K=3$. Mean CERs over 50 simulated data sets are reported, with
standard errors in parentheses. The Poisson-based measures perform
quite well when overdispersion is low, but tend to be outperformed by
EdgeR in the presence of substantial overdispersion}
\label{tab:simclust}
\begin{tabular*}{\tablewidth}{@{\extracolsep{\fill}}ld{1.2}cc@{}}
\hline
$\bolds{\phi}$&\multicolumn{1}{c}{$\bolds{\sigma}$}
& \multicolumn{1}{c}{\textbf{Method}}
& \multicolumn{1}{c@{}}{\textbf{Clustering error rate}}\\
\hline
0.01 & 0.15 &Cai & 0.3592 (0.0071)\\
&&Berninger & 0.5704 (0.0173)\\
&&EdgeR & 0.0000 (0.0000)\\
&&VST & 0.6201 (0.0029)\\
&&Squared Euclidean total count & 0.5675 (0.0191)\\
&&Squared Euclidean quantile & 0.5662 (0.0215)\\
&&Squared Euclidean median ratio & 0.5755 (0.0178)\\
&&Poisson total count & 0.0045 (0.0045)\\
&&Poisson quantile & 0.0057 (0.0047)\\
&&Poisson median ratio & 0.0045 (0.0045)\\
[6pt]
0.1 & 0.2 & Cai & 0.3803 (0.0058)\\
&&Berninger & 0.1905 (0.0258)\\
&&EdgeR & 0.0000 (0.0000)\\
&&VST & 0.6204 (0.0029)\\
&&Squared Euclidean total count & 0.3051 (0.0327)\\
&&Squared Euclidean quantile & 0.2875 (0.0325)\\
&&Squared Euclidean median ratio & 0.3297 (0.0350)\\
&&Poisson total count & 0.2053 (0.0225)\\
&&Poisson quantile & 0.2067 (0.0228)\\
&&Poisson median ratio & 0.2006 (0.0219)\\
[6pt]
1 & 0.5 & Cai & 0.3797 (0.0063)\\
&&Berninger & 0.5309 (0.0143)\\
&&EdgeR & 0.0098 (0.0054)\\
&&VST & 0.6058 (0.0089)\\
&&Squared Euclidean total count & 0.1630 (0.0242)\\
&&Squared Euclidean quantile & 0.2190 (0.0235)\\
&&Squared Euclidean median ratio & 0.1998 (0.0305)\\
&&Poisson total count & 0.2699 (0.0255)\\
&&Poisson quantile & 0.2699 (0.0255)\\
&&Poisson median ratio & 0.2749 (0.0254)\\
\hline
\end{tabular*}
\end{table}

Interestingly, the choice
of size factor estimate (total count, quantile, or median ratio) seems
to have little effect on the classifiers' performances,
despite the fact that the choice of estimate has been found to play a
critical role in the detection of differentially expressed features
[\citet{Bullard2010}, \citet{RobinsonOshlack2010},
\citet{AndersHuber10}].
This is likely because the size factor estimates yield very different
results primarily in the setting where\vadjust{\goodbreak}
a subset of the features containing a large proportion of the counts
are highly differentially expressed. However, in such a setting,
classification tends to be quite easy. In more challenging classification
settings in which differentially expressed features have fewer counts and
display smaller differences between classes, such as in Table \ref
{tab:simclass}, the effect of the size factor estimate appears to play
a less important role.\vadjust{\goodbreak}

\subsection{Evaluation of Poisson clustering}
\label{simstudy.clust}
We compare the performances of ten clustering proposals: the $K$-means
clustering proposal of \citet{Cai04} which assumes a Poisson
model, as well as complete-linkage hierarchical clustering applied
to the nine dissimilarity measures described in Section \ref
{sec:cluster}. \citeauthor{Cai04}'s (\citeyear{Cai04}) proposal was performed with $K=3$
(the true number of clusters),
and the hierarchical clustering dendrograms for the other methods were
cut at a height that resulted
in three clusters.

To evaluate the performances of these clustering methods, we use the
clustering error rate (CER), which measures the extent to which two
partitions $P$ and $Q$ of a
set of $n$ observations disagree. Let $1_{P(i,i')}$ be an indicator for
whether observations $i$ and $i'$ are in the same group in partition
$P$, and define $1_{Q(i,i')}$ analogously. Then CER
is defined as
%
\begin{equation}
\sum_{i>i'} \bigl|1_{P(i,i')} - 1_{Q(i,i')}\bigr|\Big/\pmatrix{n \cr 2}.
\end{equation}
This is also one minus the Rand Index [\citet{Rand71}]. We took
$P$ to be the true class labels and $Q$ to be the class labels
estimated via clustering;
a~small value indicates an accurate clustering.

Simulation study results with $n=25$ observations are shown in Table~\ref{tab:simclust}. The Poisson clustering proposed in this paper
performs well for the full range of overdispersion parameters
considered. This is in part because the transformation described in
Section \ref{overdisp} makes the data
approximately Poisson even when the overdispersion parameter $\phi$ is
large. Even though the
method of \citet{Zavolan08} is based on a model that is very
similar to ours, it exhibits worse performance.
This is likely due to numerical issues with their proposal whereby two
different observations can have zero dissimilarity and two identical
observations can have nonzero dissimilarity.
It is difficult to compare the Poisson-based proposal of \citet
{Cai04} directly to the other nine since it uses a $K$-means approach,
where the number of clusters must be specified in advance. Moreover,
the proposals of \citet{Zavolan08}
and \citet{Cai04} do not entail first performing a power
transformation on the data.

In
Table \ref{tab:simclust} the EdgeR dissimilarity measure exhibited
essentially the same performance as our Poisson clustering measure when
$\phi=0.01$, and better
performance in the presence of moderate or severe overdispersion.
However, it is quite computationally intensive. The example shown in
Table \ref{tab:simclust} contains only 25 observations because running
EdgeR using the \texttt{Bioconductor} package provided by the authors
[\citet{edgeR}] is too slow for larger values of $n$. For instance,
on a~simulated example with $n=50$ and $p=10\mbox{,}000$, it took 6 minutes to
compute the dissimilarity matrix
on a~AMD Opteron 848 2.20 GHz processor. In contrast, computing
the Poisson dissimilarity matrix on the same example took 14 seconds.

In summary, the Poisson dissimilarity measure outperforms all of the
methods besides EdgeR. EdgeR is the overall winner, but is much more
computationally demanding.

\section{Application to sequencing data sets}
\label{realstudy}

\subsection{Data sets}
\label{realstudy.data}
We present results based on four data sets. The first three are RNA
sequencing data sets, and the fourth is a chromatin immunoprecipitation
(ChIP) sequencing data set
intended as a preliminary assessment of the extent to which the methods
proposed here can be applied to other types of sequencing data.

\begin{longlist}[]
\item[\textit{Liver and kidney.}] An RNA sequencing data set quantifying
the expression of 22,925 genes [\citet{Marioni08}]. There are
seven technical replicates from a liver sample and seven technical
replicates from a kidney sample, each from a single human male.
The liver and kidney samples are treated as two separate classes. The
data are available as a Supplementary File associated with \citet
{Marioni08}.
\item[\textit{Yeast.}] An RNA sequencing data set consisting of replicates
of \textit{Saccharomyces cerevisiae} (yeast) cultures [\citet
{Nagalakshmi2008}].
Three replicates were obtained for each of two library preparation
protocols, ``random
hexamer'' (RH) and ``oligo(dT)'' (dT). For each library preparation
protocol, there is an ``original'' replicate, a ``technical'' replicate
of that original replicate, and also a ``biological''
replicate. The number of reads mapping to each of 6,874 genes is
available as a Supplementary
File associated with \citet{AndersHuber10}.
In the analysis that follows, we treat the RH and dT library
preparations as two distinct classes.
\item[\textit{Cervical cancer.}] An RNA sequencing data set quantifying the
expression of microRNAs in tumor and nontumor human cervical tissue
samples [\citet{WittenFire2010}].
MicroRNAs are small RNAs, 18--30 nucleotides in length, that have been
shown to play an important regulatory role in a number of biological processes.
The data take the form of 29 tumor and 29 nontumor cervical tissue
samples with measurements on 714 microRNAs. Of the tumor samples, 6 are
adenocarcinomas (ADC), 21 are squamous cell carcinomas (SCC), and 2 are
unclassified. The two
unclassified samples are excluded from the analysis. Normal, ADC, and
SCC are treated as three separate classes.
The data are available from Gene Expression Omnibus [\citet{GEO}]
under accession number GSE20592.
Since this is a~small RNA data set,
the experimental protocol differs slightly from the description in
Section \ref{introsec}:
small RNAs were isolated before being converted to cDNA, which was then
amplified and sequenced.
As pointed out by a~reviewer, \citet{Linsen09} found that small
RNA digital gene expression profiling is biased
toward certain small RNAs, and so small RNA sequencing data sets cannot
be used to accurately determine absolute numbers of small RNAs.
However, this\vadjust{\goodbreak} bias is systematic and reproducible, and so small RNA
sequencing data sets can be used to determine relative
expression differences between samples. The classification and
clustering proposals in this paper rely on relative rather than
absolute expression differences in the sense
that accurate classification or clustering can be performed even if
certain small RNAs contain a disproportionately large number of counts
relative to the true abundance in the original sample.
\item[\textit{Transcription factor binding.}] ChIP sequencing is a new
approach for mapping protein-DNA interactions at a genome-wide level
that relies
upon recently developed techniques for high throughput DNA sequencing
[\citet{Johnson07}]. Like RNA sequencing, the results of a ChIP
sequencing experiment can be arranged as a $n \times p$ matrix with $n$
observations and $p$
features. The features represent the DNA binding regions for a protein
of interest, and the $(i,j)$ element of the data matrix indicates
the number of times that the protein was observed to bind to the
$j$th binding region in the $i$th sample.
In \citet{Kasowski2010}, the binding sites of RNA polymerase II
were mapped in each of ten individuals. 19,061 binding regions were
identified, each of which was treated
as a distinct feature.
At least three replicates were available for each individual, and there
were 39 observations in total. This data are available as a Supplementary
File associated with \citet{AndersHuber10}.
In what follows, we treat each of the ten individuals as a distinct class.
\end{longlist}

\subsection{Evaluation of sparse PLDA}
\label{realstudy.class}
\begin{figure}

\includegraphics{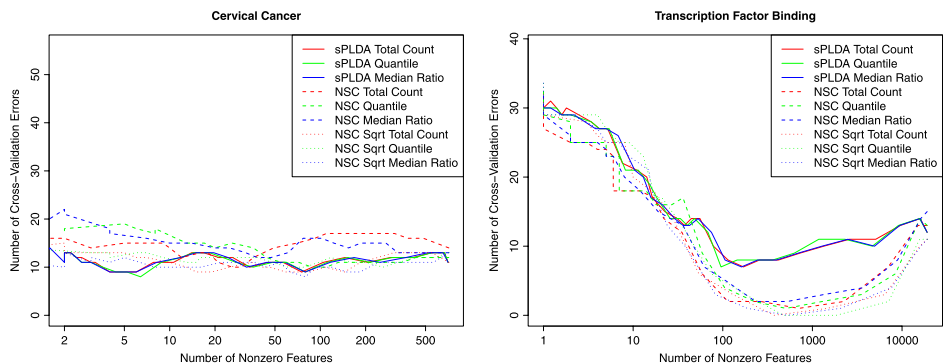}%
\vspace*{-2pt}
\caption{A comparison of classification methods on the cervical cancer
data of Witten et~al. (\protect\citeyear{WittenFire2010}) and the
transcription factor binding data of Kasowski et~al.
(\protect\citeyear{Kasowski2010}). NSC, NSC on the square root
transformed data, and sPLDA were performed each with three distinct
size factor estimates---total count, quantile [Bullard et~al.
(\protect\citeyear{Bullard2010})], and median ratio [Anders and Huber
(\protect\citeyear{AndersHuber10})], each of which is described in
Section \protect\ref{sectionmod}. Five-fold cross-validation was
performed. The resulting cross-validation error curves are shown as a
function of the number of features included in the classifier. Results
for the yeast data of Nagalakshmi et~al.
(\protect\citeyear{Nagalakshmi2008}) and the liver and kidney data of
Marioni et~al. (\protect\citeyear{Marioni08}) are not shown because all
methods gave 0 cross-validation errors for all of the tuning parameter
values considered.}\label{realclass}
\vspace*{-6pt}
\end{figure}

A total of nine classification methods were compared: NSC, NSC on
$\sqrt{X_{ij}+3/8}$, and sPLDA, each with three different size factor
estimates. Details are given in Section
\ref{simstudy.class}. These methods were applied to the four data sets
described in Section \ref{realstudy.data}.
Results on the cervical cancer and transcription factor binding data
sets are shown in Figure \ref{realclass}. Results for the liver and
%
kidney data and the yeast data are not shown since on those two data
sets, all methods gave 0 cross-validation errors for all of the tuning
parameter values considered.

\subsection{Evaluation of Poisson clustering}
\label{realstudy.clust}

\begin{figure}
\includegraphics{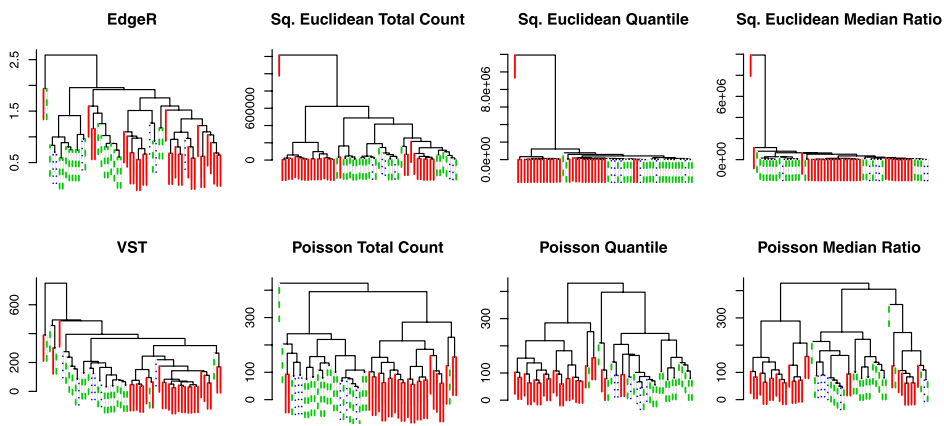}%
\vspace*{-2pt}
\caption{Complete-linkage hierarchical clustering dendrograms obtained
by computing eight
dissimilarity measures on the cervical cancer data of
Witten et~al. (\protect\citeyear{WittenFire2010}). The dissimilarity measures are described
in Table \protect\ref{tab:cluster}.
Normal samples are shown as red solid lines, ADC as blue dotted lines,
and SCC as green dashed lines.
The Poisson-based approaches seem to do the best job of separating the
normal samples from the tumor samples.}\label{fireclust}
\vspace*{-3pt}
\end{figure}

\begin{figure}

\includegraphics{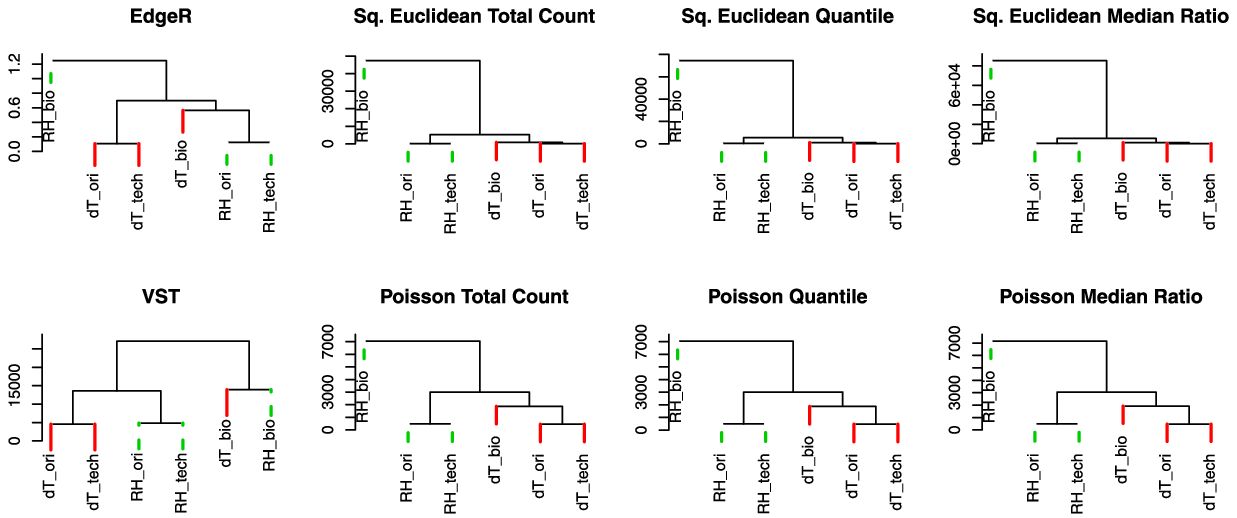}%
\vspace*{-2pt}
\caption{Complete-linkage hierarchical clustering dendrograms obtained
by computing eight dissimilarity
measures on the yeast data of Nagalakshmi et~al.
(\protect\citeyear{Nagalakshmi2008}).
The dissimilarity measures are described
in Table \protect\ref{tab:cluster}.
The dT samples are shown in
red and RH samples are in green. The three methods based on Euclidean
distance and the three methods based on Poisson dissimilarity give the
most accurate dendrograms on this data, since they successfully group
the dT samples together.}\label{nagaclust}
\vspace*{-6pt}
\end{figure}

\begin{figure}
\includegraphics{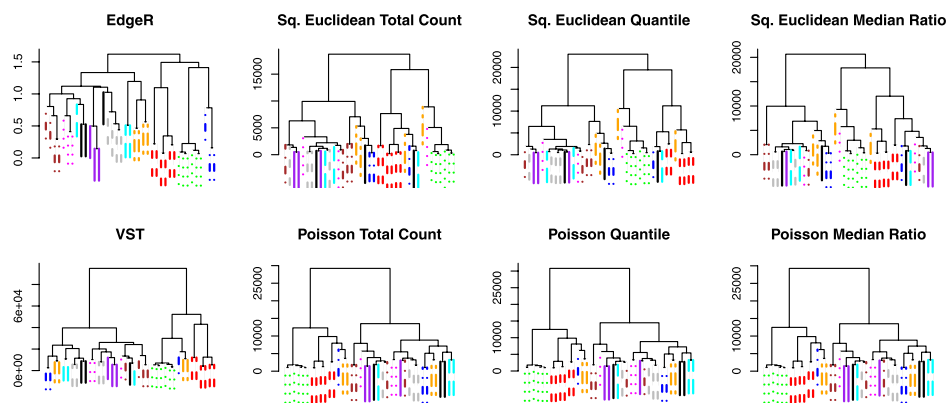}%
\vspace*{-2pt}
\caption{Complete-linkage hierarchical clustering dendrograms obtained
by computing eight dissimilarity measures on the
transcription factor binding data of Kasowski et~al.
(\protect\citeyear{Kasowski2010}). The dissimilarity measures are described in
Table \protect\ref{tab:cluster}. Replicates from each of the ten
individuals are shown in a different color.}\label{kasclust}
\end{figure}

\begin{figure}

\includegraphics{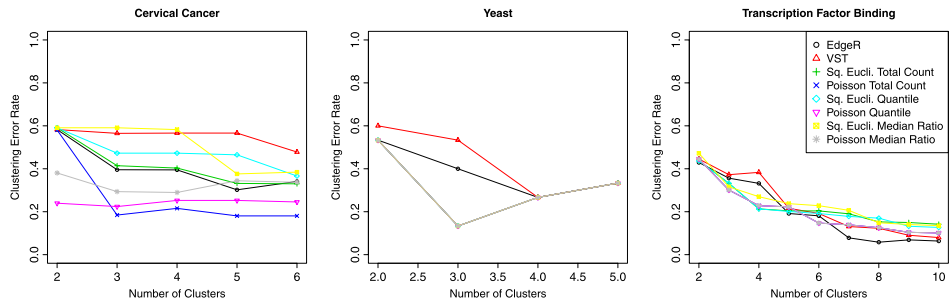}%
\caption{For each of the eight dissimilarity measures described in
Table \protect\ref{tab:cluster}, the figure displays
the CERs that result from cutting the complete linkage hierarchical
clustering dendrogram at various cutpoints.}\label{cutree}
\end{figure}

We clustered the observations in each of the four data sets described
in Section \ref{realstudy.data}. Eight dissimilarity measures were
used to perform complete-linkage hierarchical clustering
(Table~\ref{tab:cluster}).
The liver and kidney data resulted in a
perfect clustering by all methods of comparison (results not shown).
The
cervical cancer, yeast, and transcription factor binding results are
shown in Figures \ref{fireclust}, \ref{nagaclust} and \ref{kasclust}.
%
The cervical cancer data are challenging: the Poisson dissimilarity
measures are best able to distinguish between tumor and nontumor
samples, but
no method is able to convincingly distinguish between ADC and SCC
(Figure \ref{fireclust}). For the yeast data, all methods but EdgeR
and VST yield essentially the same dendrogram---one RH sample appears
to be distinct
%
from all the other samples, but the remaining RH and dT samples are
quite distinct. For that data, EdgeR and\vadjust{\goodbreak} VST yield different (and
presumably worse) clusterings.
%
The transcription factor binding data
are the most complex since there are ten\vadjust{\goodbreak} groups (one per individual).
There is no clear winner, but EdgeR seems to perform quite well (Figure
\ref{kasclust}). The CERs for each dendrogram can be found in Figure
\ref{cutree}.

\section{Discussion}
\label{disc}
In this paper we have proposed approaches for the classification and
clustering of sequencing data. As sequencing
technologies become increasingly widespread, the importance of
statistical methods that are well suited to this type of data will
increase. The approaches proposed in this paper were designed for
RNA sequencing data, but could likely be extended to other sequencing
technologies such as DNA and ChIP
sequencing. In fact, an application to ChIP sequencing data was
presented in Section \ref{realstudy}.

%

The methods proposed in this paper follow naturally from a
simple Poisson log linear model (\ref{poisson.classes})
for sequencing data. Similar approaches could be taken using an
alternative model, for instance, one based on the
negative binomial distribution. The methods proposed seem to work very
well if the true model for the data is Poisson or if there is mild
overdispersion relative to the Poisson model.
Performance degrades in the presence of severe overdispersion. Most sequencing
data seem to be somewhat overdispersed relative to the Poisson model.
It may be that extending the approaches proposed here to the negative
binomial model could result in improved performance in the
presence of overdispersion.

A number of authors have proposed detecting
differentially expressed features in sequencing data by making
inference on a Poisson log
linear model
[see, e.g., \citet{Marioni08}, \citet{Bullard2010},
\citet{JunSamSeq}]. In this
paper we have used such a model to develop proposals for
classification and clustering. However,
many other types of inference based on sequencing data are likely to
be of interest in the future. For instance, in a recent
paper, \citet{Sparsebinarypca} proposed a method for performing
principal components analysis
(PCA) for high-dimensional binary data. In a similar vein, one could
develop an approach for
PCA for sequencing data using the Poisson log linear model (\ref{mostbasic}).
We leave this as a topic for further
research.

It has been shown that the manner in which samples are normalized is of
great importance in identifying differentially expressed
features on the basis of sequencing data [\citet{Bullard2010},
\citet{RobinsonOshlack2010}, \citet{AndersHuber10}].
However, in Sections \ref{simstudy} and \ref{realstudy}, the
normalization approach
appeared to have little effect on the results obtained. This seems to
be due to the fact that the choice of normalization approach is most important
when a few features with very high counts are differentially expressed
between classes. In that case, identification of differentially
expressed features can be
challenging, but classification and clustering are quite straightforward.

It is known that RNA-sequencing data suffer from \textit{transcript
length bias}---that is,\vadjust{\goodbreak} longer transcripts tend to
result in a greater number of reads, resulting in an increased tendency
to call such transcripts differentially expressed
[\citet{OshlackWakefield09}].
In a similar manner, the classification and clustering proposals made
in this paper are affected by the total number of counts per feature;
this can be seen by inspection of (\ref{postmean}), (\ref
{sparseest}) and (\ref{loglrpoi}).
It seems clear that bias due to the total number of counts per feature
is undesirable for the task of identifying differentially expressed
transcripts, since it makes it difficult to detect differential
expression for low-frequency transcripts.
However, it is not clear that such a bias is undesirable in the case of
classification or clustering, since we would like
features about which we have the most information---namely, the
features with the highest total counts---to have the greatest effect
on the classifiers and dissimilarity measures that we use. More
investigation into this matter is left as a topic for future
work.\looseness=-1

Our proposal for clustering sequencing data is based on the
development of a dissimilarity measure that is potentially more
appropriate for count data than
standard Euclidean distance. The resulting
dissimilarity matrix can then be input to a standard clustering
algorithm, such as hierarchical clustering. Other statistical
techniques that rely upon a dissimilarity matrix, such as
multidimensional scaling,
could also be performed using the Poisson dissimilarity measure
developed here.

An \texttt{R} language package implementing the methods proposed in this
paper will be
made available.

\begin{appendix}\label{app}
\section{Properties of the Poisson dissimilarity~measure}
We wish to prove that the dissimilarity between ${\mathbf x}_i$ and ${\mathbf x}_{i'}$ specified in~(\ref{loglrpoi}) is nonnegative, and that it
equals zero when ${\mathbf x}_i = {\mathbf x}_{i'}$.

To prove nonnegativity, first notice that $g(d_{ij}) = - \hat N_{ij}
d_{ij} + X_{ij} \log d_{ij}$ is a concave function of $d_{ij}$, and is
maximized when $d_{ij} = \frac{X_{ij}}{\hat N_{ij}}$. Therefore,
$g(\frac{X_{ij}}{\hat N_{ij}}) \geq g(1)$. And since $\frac{X_{ij} +
\beta}{\hat N_{ij} + \beta}$ is between 1 and $\frac{X_{ij}}{\hat
N_{ij}}$, concavity of $g$ ensures that $g(\frac{X_{ij} + \beta}{\hat
N_{ij} + \beta}) \geq g(1)$; that\vspace*{2pt} is, $\hat N_{ij} - \hat N_{ij}
d_{ij} + X_{ij} \log d_{ij} \geq0$. It follows directly that (\ref
{loglrpoi}) is nonnegative.

Now, if ${\mathbf x}_i = {\mathbf x}_{i'}$, then $\hat N_{ij} = \hat N_{i'j} =
X_{ij} = X_{i'j}$ and so by (\ref{part2}), $\hat d_{ij}= \hat
d_{i'j}=1$. By inspection, (\ref{loglrpoi}) equals zero.

\section{Equivalence of log likelihood ratio statistics under Poisson
model and multinomial model}
Here, we show that the log likelihood ratio statistic (\ref{loglrpoi})
under the Poisson model is identical to the log likelihood ratio
statistic under the model of \citet{Zavolan08}, if
appropriate estimates of $N_{ij}$, $N_{i'j}$, $d_{ij}$, and $d_{i'j}$
are used in (\ref{loglrpoi}).

In Section \ref{proposal.clust.deets} we assumed that the $i$th and
$i'$th observations take the form
$X_{ij} \sim\Poisson(N_{ij} d_{ij}), X_{i'j} \sim\Poisson(N_{i'j}
d_{i'j}), N_{ij} = s_i g_j, N_{i'j} = s_{i'} g_j$. Under the null
hypothesis, $d_{ij} = d_{i'j}=1$. Under the alternative,
$d_{ij}$ and $d_{i'j}$ are unconstrained.
Suppose we estimate $N_{ij}$ and $N_{i'j}$ under the null using the
MLEs, or, equivalently, using the total count size factor estimates
given in Section \ref{sectionmod}: then
$\hat N_{ij} = X_{i \cdot} (X_{ij} + X_{i'j})/(X_{i \cdot} + X_{i'
\cdot})$, $\hat N_{i'j} = X_{i' \cdot} (X_{ij} + X_{i'j})/(X_{i \cdot
} + X_{i' \cdot})$.
Treating these estimates as offsets under the alternative,
the MLE for $d_{ij}$ is $X_{ij}/N_{ij}$, and the MLE for $d_{i'j}$ is
$X_{ij}/N_{i'j}$. Plugging these estimates into (\ref{loglrpoi}) yields
%
\begin{eqnarray}\label{loglr.writtenout}
&& \sum_{j=1}^p \bigl( X_{i \cdot} (X_{ij} + X_{i'j})/(X_{i \cdot} +
X_{i' \cdot}) + X_{i' \cdot} (X_{ij} + X_{i'j})/(X_{i \cdot} + X_{i'
\cdot})\nonumber\\
&&\hspace*{31pt}\quad{} - X_{ij} - X_{i'j}
+ X_{ij} \log(X_{ij}/N_{ij}) + X_{i'j} \log(X_{i'j}/N_{i'j})\bigr)
\nonumber\\
&&\qquad= \sum_{j=1}^p \bigl( X_{ij} \log(X_{ij}/N_{ij}) + X_{i'j} \log
(X_{i'j}/N_{i'j}) \bigr) \\
&&\qquad= \sum_{j=1}^p \bigl( X_{ij} \log X_{ij} + X_{i'j} \log X_{i'j} - (X_{ij}
+ X_{i'j}) \log(X_{ij} + X_{i'j}) \bigr)\nonumber\\
&&\qquad\quad{} + (X_{i \cdot} + X_{i' \cdot})
\log(X_{i \cdot} + X_{i' \cdot})
- X_{i \cdot} \log X_{i \cdot} - X_{i' \cdot} \log X_{i'
\cdot}.\nonumber
\end{eqnarray}

Now, \citet{Zavolan08} instead assume a multinomial model for the
data:
$X_{i1}, \ldots, X_{ip} \sim\Multinomial(X_{i \cdot}, q_1, \ldots, q_p)$
and $X_{i'1}, \ldots, X_{i'p} \sim\break\Multinomial(X_{i' \cdot}, r_1,
\ldots, r_p)$. Under the null, $q_j = r_j$ $\forall j$. Under the
alternative, $q_j$ and $r_j$ are unconstrained. This results
in the likelihood ratio statistic
%
\begin{equation}\label{lrmult}
\prod_{j=1}^p\frac{ (X_{ij}/X_{i \cdot})^{X_{ij}} (X_{i'j}/X_{i'
\cdot})^{X_{i'j}}} {((X_{ij} + X_{i'j})/(X_{i \cdot} + X_{i' \cdot
}))^{X_{ij} + X_{i'j}}}.
\end{equation}
Taking the logarithm of (\ref{lrmult}) yields (\ref{loglr.writtenout}).

Note that, in practice, the dissimilarity measures proposed in Section~%
\ref{proposal.clust.deets} and in \citet{Zavolan08} are not
identical, since in Section \ref{proposal.clust.deets}
we estimate $d_{ij}$
and $d_{i'j}$ as the posterior means using a Gamma prior. \citet
{Zavolan08} instead use a Dirichlet prior on $q_1, \ldots, q_p$ and
$r_1, \ldots, r_p$
and use the Bayes factor as the dissimilarity measure.
In fact, the proposal of \citet{Zavolan08} seems to perform substantially
worse than that of Section \ref{proposal.clust.deets} in the
simulation study in Section \ref{simstudy}.
\end{appendix}

%

\section*{Acknowledgments}

The author thanks Robert Tibshirani and Jun Li at Stanford University
for helpful conversations. Also, thanks to Mihaela Zavolan for
providing software for a proposal described in
\citet{Zavolan08}, to Li Cai for providing command line
software for the proposal described in \citet{Cai04}, and
to Mark Robinson for helpful responses to inquiries regarding the
Bioconductor package edgeR [\citet{edgeR}]. An Associate
Editor and two referees provided comments that greatly improved the
quality of this paper.


%

%
\printaddresses

\end{document}